\documentclass[submission,copyright,creativecommons]{eptcs}
\providecommand{\event}{Fifteenth International Workshop on Graph\\Computation Models (GCM 2024)} 

\newcommand{\gp}{\texorpdfstring{GP\,2}{}}

\usepackage{iftex}
\usepackage{algpseudocode}
\usepackage{graphicx}
\usepackage{amsmath}
\usepackage[table,xcdraw]{xcolor}
\usepackage{pgf, tikz}
\usepackage{algorithm}
\usepackage{fancyvrb}
\usepackage{mdframed}
\usepackage{pgfplotstable}
\usepackage{pgfplots}
\usepackage{capt-of}

\definecolor{plot1}{RGB}{70, 116, 193}
\definecolor{plot2}{RGB}{235, 125, 60}
\definecolor{plot3}{RGB}{165, 165, 165}
\definecolor{plot4}{RGB}{252, 190, 45}
\definecolor{plot5}{RGB}{94, 156, 210}
\definecolor{plot6}{RGB}{113, 171, 77}
\definecolor{plot7}{RGB}{156, 72, 25}
\definecolor{plot8}{RGB}{40, 69, 117}
\definecolor{gp2grey}{RGB}{150, 150, 150}

\pgfplotsset{compat=1.16}

\usepackage{iftex}

\ifpdf
  \usepackage{underscore}         
  \usepackage[T1]{fontenc}        
\else
  \usepackage{breakurl}           
\fi

\title{Linear-Time Graph Programs without Preconditions}
\author{
Ziad Ismaili Alaoui
\thanks{This author's work was done while he was affiliated with the University of York.}
\institute{Department of Computer Science, University of Liverpool \\ Liverpool, United Kingdom}
\email{ziad.ismaili-alaoui@liverpool.ac.uk}
\and
Detlef Plump
\institute{Department of Computer Science, University of York \\ York, United Kingdom}
\email{detlef.plump@york.ac.uk}
}
\def\titlerunning{Linear-Time Graph Programs without Preconditions}
\def\authorrunning{Z. Ismaili Alaoui \& D. Plump}
\begin{document}
\maketitle

\begin{abstract}
  We report on a recent breakthrough in rule-based graph programming, which allows us to reach the time complexity of imperative linear-time algorithms. In general, achieving the complexity of graph algorithms in conventional languages using graph transformation rules is challenging due to the cost of graph matching. Previous work demonstrated that with \emph{rooted} rules, certain algorithms can be executed in linear time using the graph programming language \gp{}. However, for non-destructive algorithms that retain the structure of input graphs, achieving linear runtime required input graphs to be connected and of bounded node degree. In this paper, we overcome these preconditions by enhancing the graph data structure generated by the \gp{} compiler and exploiting the new structure in programs. We present three case studies, a cycle detection program, a program for numbering the connected components of a graph, and a breadth-first search program. Each of these programs runs in linear time on both connected and disconnected input graphs with arbitrary node degrees. We give empirical evidence for the linear time complexity by using timings for various classes of input graphs.
\end{abstract}

\section{Introduction}
Designing and implementing languages for rule-based graph rewriting, such as GReAT \cite{Agrawal-Karsai-Neema-Shi-Vizhanyo06a}, GROOVE \cite{Ghamarian-deMol-Rensink-Zambon-Zimakova12a}, GrGen.Net \cite{Jakumeit-Buchwald-Kroll10a}, Henshin \cite{Struber-Born-Gill-Groner-Kehrer-Ohrndorf-Tichy17a}, and PORGY \cite{Fernandez-Kirchner-Pinaud19a}, poses significant performance challenges. Typically, programs written in these languages do not achieve the same runtime efficiency as those written in conventional imperative languages such as C or Java. The primary obstacle is the cost of graph matching, where matching the left-hand graph $L$ of a rule within a host graph $G$ generally requires time $|G|^{|L|}$, with $|X|$ denoting the size of graph $X$. (Since $L$ is fixed, this is a polynomial.) As a consequence, standard imperative graph algorithms running in linear time (see, for example, \cite{Cormen-Leiserson-Rivest-Stein22a,Skiena20a}) may exhibit non-linear, polynomial runtimes when recast as rule-based graph programs.

To address this issue, the graph programming language \gp{} \cite{plump2012design} supports \emph{rooted} graph transformation rules, initially proposed by D\"orr \cite{dorr1995efficient}. This approach involves designating certain nodes as \emph{roots} and matching them with roots in the host graphs. Consequently, only the neighbourhoods of host graph roots need to be searched for matches, which can often be done in constant time under mild conditions. The \gp{} compiler \cite{bak2015gp} maintains a list of pointers to roots in the host graph, facilitating constant-time access to roots if their number remains bounded throughout the program's execution. In \cite{bak2012rooted},\, \emph{fast} rules were identified as a class of rooted rules that can be applied in constant time, provided host graphs contain a bounded number of roots and have a bounded node degree.

The first linear-time graph problem implemented by a GP\,2 program was 2-colouring. In \cite{bak2012rooted,bak2015gp}, it is shown that this program colours connected graphs of bounded node degree in linear time. Since then, the \gp{} compiler has received some major improvements, particularly related to the runtime graph data structure used by the compiled programs \cite{campbell2020improved}. These improvements made a linear-time worst-case performance possible for a wider class of programs, in some cases even on input graph classes of unbounded degree. See \cite{campbell2022fast} for an overview.

Despite this progress, programs that retain the structure of input graphs, such as the aforementioned 2-colouring program, have until now required non-linear runtimes on disconnected graphs. The problem is that, after a connected component is visited, the number of failed attempts to match a non-visited node in a different connected component may increase. Consequently, in disconnected graph classes, this number may grow quadratically in the graph size, leading to a quadratic program runtime. In connected graph classes, this undesirable behaviour is ruled out because all nodes are reachable from a single undirected depth-first search.

In this paper, we present two updates to the \gp{} compiler, one being introduced in \cite{ismaili2024linear}, which allow lifting the preconditions that host graphs must be connected and have a bounded node degree. In short, the solution is to improve the graph data structure generated by the compiler. Nodes are now stored in separate linked lists based on their \emph{marks} (red, green, blue, grey or unmarked), and each node comes with a two-dimensional array of linked lists storing all incident edges based on their marks (red, green, blue, dashed or unmarked) and orientation (incoming, outgoing or looping). This enables the matching algorithm to find in constant time a node with a specific mark or an edge with a specific mark and orientation. For instance, if a red node is needed, a single access to the list of red nodes will either locate such a node or confirm its absence. Similarly, if an outgoing green edge is required, a single access to the corresponding linked list will either find such an edge or determine that none exists.

In addition to the new graph data structure, a programming technique is needed to take advantage of the improved storage. In a case study, we demonstrate how to recognize acyclic graphs in linear time with the new graph representation. Our program admits both connected and disconnected input graphs with arbitrary node degrees. It either detects that a graph is cyclic or returns an acyclic graph that is isomorphic to the input graph up to marking. Then, we discuss two more programs relying on the improved compiler: one numbering the connected components of a graph, and another performing a breadth-first search. For both programs, we give empirical evidence that they run in linear time on different classes of input graphs.


\section{The Problem with Graph Classes of Unbounded Degree}
\label{s:unbounded}
The current section and the next explain the problem caused by classes of input graphs with an unbounded node degree, and how the updated compiler overcomes this problem. We include this material from \cite{ismaili2024linear} for the benefit of the reader. For a description of the GP\,2 programming language, we refer to \cite{campbell2022fast}. 

\begin{figure}[!ht]
\begin{mdframed}[linewidth=0.8pt]
\begin{verbatim}
Main  = try init then (DFS!; Check)
DFS   = forward!; try back else break
Check = if match then fail

\end{verbatim}

\begin{tikzpicture}
\tikzstyle{every node}=[font=\ttfamily]

\draw (0.9,0.75) node[align=left] {init(x:list)};
\node[rectangle, thick, thick, rounded corners=7, draw=black, minimum size=5mm, fill=gray!50, label=below:\tiny\tiny 1](a2) at (0,0){x};
\draw (1,0) node[] {$\Rightarrow$};
\node[rectangle, thick, rounded corners=7, draw=black, minimum size=5mm,  double, double distance=1pt, fill=cyan, label=below:\tiny\tiny1](a2) at (2,0){x};
\draw[->, thick];
\draw (4.8,1) -- (4.8,-0.5);
\end{tikzpicture}
\begin{tikzpicture}
\tikzstyle{every node}=[font=\ttfamily]
\hspace{1.55em}
\draw (1,0.75) node[align=left] {match(x:list)};
\node[rectangle, thick, rounded corners=7, draw=black, minimum size=5mm, fill=gray!50, label=below:\tiny\tiny1](a2) at (0,0){x};
\draw (1,0) node[] {$\Rightarrow$};
\node[rectangle, thick, rounded corners=7, draw=black, minimum size=5mm, fill=gray!50, label=below:\tiny\tiny 1](a2) at (2,0){x};
\draw[->, thick];

\end{tikzpicture}

\begin{tikzpicture}
\tikzstyle{every node}=[font=\ttfamily]
\draw (1.6,0.75) node[align=left] {forward(x,y,z:list)};
\node[rectangle, thick, rounded corners=7, draw=black, minimum size=5mm,  double, double distance=1pt, fill=cyan, label=below:\tiny\tiny1](a1) at (0,0){x};
\node[rectangle, thick, rounded corners=7, draw=black, minimum size=5mm, fill=gray!50, label=below:\tiny\tiny2](a2) at (1,0){y};
\draw (2,0) node[] {$\Rightarrow$};
\node[rectangle, thick, rounded corners=7, draw=black, minimum size=5mm, fill=cyan, label=below:\tiny\tiny1](a3) at (3,0){x};
\node[rectangle, thick, rounded corners=7, draw=black, minimum size=5mm, double, double distance=1pt, fill=cyan, label=below:\tiny\tiny2](a4) at (4,0){y};
\draw[-, line width=1.2pt] (a1) edge[] node[above, color = black]{z} (a2) (a3) edge[dashed] node[above, color = black]{z} (a4);
\draw (4.8,1) -- (4.8,-0.5);
\end{tikzpicture}
\hspace{1.5em}
\begin{tikzpicture}
\tikzstyle{every node}=[font=\ttfamily]
\draw (1.3,0.75) node[align=left] {back(x,y,z:list)};
\node[rectangle, thick, rounded corners=7, draw=black, minimum size=5mm, fill=cyan, label=below:\tiny\tiny1](a1) at (0,0){x};
\node[rectangle, thick, rounded corners=7, draw=black, minimum size=5mm, double, double distance=1pt, fill=cyan, label=below:\tiny\tiny2](a2) at (1,0){y};
\draw (2,0) node[] {$\Rightarrow$};
\node[rectangle, thick, rounded corners=7, draw=black, minimum size=5mm, double, double distance=1pt, fill=cyan, label=below:\tiny\tiny1](a3) at (3,0){x};
\node[rectangle, thick, rounded corners=7, draw=black, minimum size=5mm, label=below:\tiny\tiny2](a4) at (4,0){y};
\draw[-, line width=1.2pt] (a1) edge[dashed,shorten <=1.5] node[above, color = black]{z} (a2) (a3) edge[] node[above, color = black]{z} (a4);
\end{tikzpicture}

\end{mdframed}
\caption{The old program \texttt{is-connected}.}
\label{fig:is-con-fig-old}
\end{figure}
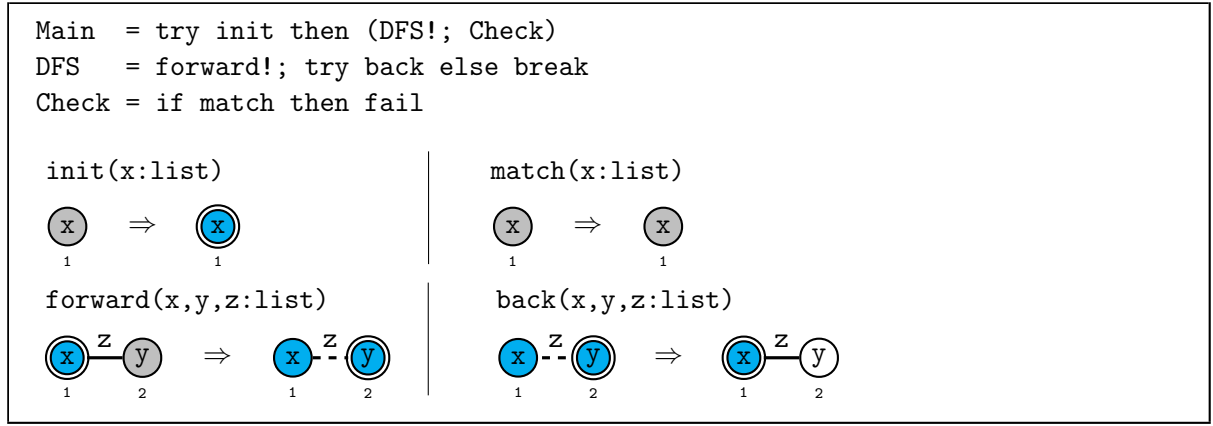

Previously, non-destructive GP\,2 programs based on depth-first-search ran in linear time on graph classes of bounded node degree but in non-linear time on graph classes of unbounded degree \cite{campbell2022fast}. For example, consider the program \texttt{is-connected} in Figure \ref{fig:is-con-fig-old} which checks whether a graph is connected.\footnote{Node labels such as \texttt{x} are written inside nodes, whereas small integers below nodes are their identifiers. Nodes without identifiers on the left-hand side are to be deleted; nodes without identifiers on the right-hand side are to be added. Nodes with the same identifier on each side are to be kept.} Input graphs are arbitrary \gp{}  host graphs with grey nodes and unmarked edges. The program fails on a graph if and only if the graph is disconnected. 

Rule \texttt{init} picks an arbitrary grey node as a root (if the input graph is non-empty) and then the loop \texttt{DFS!} performs a depth-first search of the connected component of the node chosen by \texttt{init}. The rule \texttt{forward} marks each newly visited node blue, and \texttt{back} unmarks it once it is processed. Procedure \texttt{DFS} ends when \texttt{back} fails to match, indicating that the search is complete. Rule \texttt{match} checks whether a grey-marked node still exists in the graph following the execution of \texttt{DFS!}. This is the case if and only if the input graph contains more than one connected component. In this situation the program invokes the command \texttt{fail}, otherwise it terminates by returning the graph resulting from the depth-first search.

\begin{figure}
    \centering
        \input{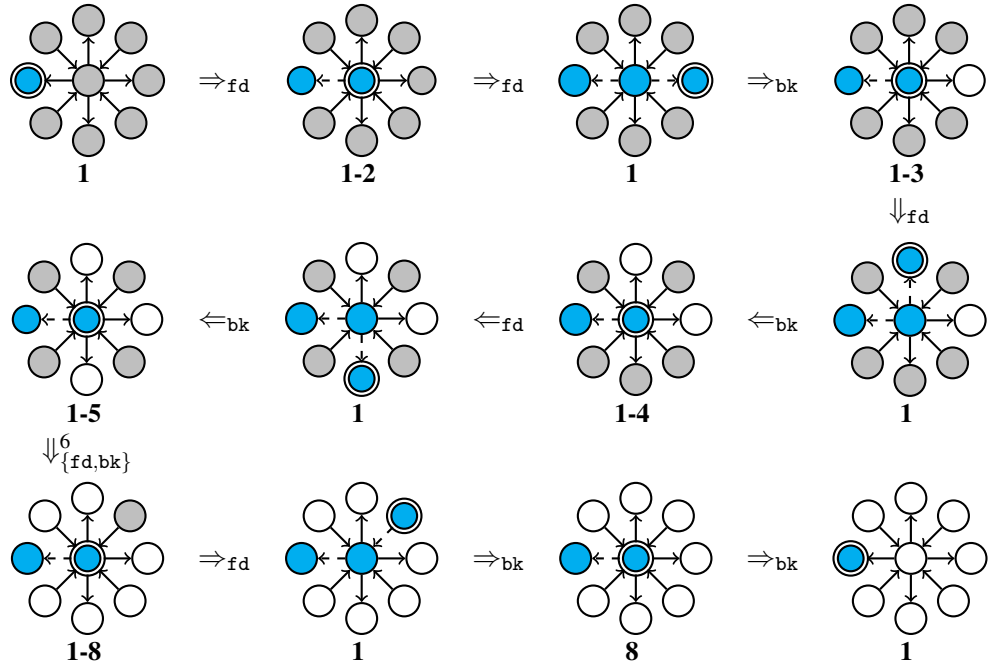}
    \caption{Matching attempts with the \texttt{forward} rule. \texttt{fd} and \texttt{bk} denote \texttt{forward} and \texttt{back}, respectively.}
    \label{fig:unbounded}
\end{figure}

It can be shown that the program \texttt{is-connected} runs in linear time on classes of graphs with bounded node degree \cite{campbell2022fast}. However, as the following example shows, the program may require non-linear time on unbounded-degree graph classes.
Figure \ref{fig:unbounded} shows an execution of \texttt{is-connected} on a star graph with $8$ edges (see also Figure \ref{fig:graph-class-2a}). The numbers below the graphs show the ranges of attempts that the matching algorithm may perform. For instance, in the second graph of the top row, either a match is found immediately among the edges that connect the central node with the grey nodes, or the dashed edge is unsuccessfully tried first. 
In order to find a match for the rule \texttt{forward}, the matching algorithm considers, in the worst case, every edge incident with the root. When the node central to the graph is rooted and the rule \texttt{forward} is called, the matching algorithm may first attempt a match with the dashed back edge and all edges incident with an unmarked node. Therefore, the maximum number of matching attempts for \texttt{forward} grows as the root moves back to the central node. As can be seen from this example, the worst-case complexity of matching \texttt{forward} throughout the program's execution is $2|E|+\sum_{i=1}^{|E|}i=\mathrm{O}(|E|^2)$ where $E$ is the set of edges.

\section{First Compiler Enhancement}
\newcommand{\ttt}{\texttt}
\newcommand{\csp}{\hspace{.2em}}
\newcommand{\bskip}{\baselineskip}
\newcommand{\red}[1]{\textcolor{red}{#1}}
\newcommand{\blue}[1]{\textcolor{blue}{#1}}
\newcommand{\green}[1]{\textcolor{green}{#1}}
\newcommand{\brown}[1]{\textcolor{brown}{#1}}
\newcommand{\black}[1]{\textcolor{black}{#1}}

To address the problem described in Section \ref{s:unbounded}, we changed the GP\,2 compiler described in \cite{campbell2020improved}, which we refer to as the \emph{2020 compiler}. We call the version introduced in this paper the \emph{new compiler}\footnote{Available at: \url{https://github.com/UoYCS-plasma/GP2}.}.
The 2020 compiler stored the host graph's structure as one linked list containing every node in the graph, with each node storing two linked lists of edges: one for incoming edges and one for outgoing edges. When iterating through edge lists to find a particular match for a rule edge, the 2020 compiler had to traverse through edges with marks incompatible with that of the rule edge. This resulted in performance issues, especially if nodes could be incident to an unbounded number of edges with marks incompatible with the edge to be matched. 

For example, consider the rule \texttt{move} from Figure \ref{fig:is-dag-fig}. Initially, the matching algorithm matches node \texttt{1} from the interface with a root node in the host graph. Subsequently, it iterates through the node's edge lists to locate a match for the red edge. In the 2020 compiler, all edges incident to this node were stored within two lists, one for each orientation, irrespective of their marks. However, if the node is incident to a growing number of unmatchable edges (because of mark changes), the matching algorithm would face, in the worst case, a growing number of iterations through the edge lists to find a single red edge.

When considering a match for a rule edge, host edges with incorrect orientation or incompatible marks do not match; thus, the matching algorithm need not iterate through them. By organising the edges incident to a node into linked lists based on their mark and orientation, the matching algorithm can selectively consider linked lists of edges of correct mark and orientation. More precisely, in the new compiler, we updated the graph structure of the 2020 compiler by replacing the two linked lists with a two-dimensional array of linked lists of edges. Each element of the array stores a linked list containing edges of a particular mark and orientation. We also consider loops to be a distinct type of orientation, separate from non-loop outgoing and incoming edges. The array, therefore, consists of $5$ rows (unmarked, dashed, red, blue, green) and $3$ columns (incoming, outgoing, loop), totalling $15$ cells that each store a single linked list. See Figure \ref{fig:enter-label} for an illustration.
\begin{figure}[htb]
    \begin{center}
    \begin{tabular}{p{15mm}|>{\centering\arraybackslash}p{8mm}|>{\centering\arraybackslash}p{8mm}|>{\centering\arraybackslash}p{8mm}|}
                & in   & out    & loop \\
  \hline
  unmarked      & \dots & \dots & \dots \\
  \hline
  dashed        & \dots & \dots & \dots \\
  \hline
  \red{red}     & \dots & \dots & \dots \\
  \hline
  \green{green} & \dots & \dots & \dots \\
  \hline
  \blue{blue}   & \dots & \dots & \dots \\
  \hline
\end{tabular} 
\end{center}
    \caption{Two-dimensional array of linked lists of edges.}
    \label{fig:enter-label}
\end{figure} 

\section{Finding Nodes in Constant Time}
\label{ss:motiv}

In this section, we explain the problem of disconnected input graphs. For example, consider the program \texttt{is-discrete} in Figure \ref{fig:is-discrete}. The program fails if and only if the input graph is discrete, that is, contains no edges. We assume that the input graph is unmarked. The program is composed of a loop followed by a test. The rule \texttt{mark} in the loop marks and roots an arbitrary unmarked node while the rule \texttt{isolated} checks whether the node rooted by \texttt{mark} is isolated. Notice that the node in the left-hand side of \texttt{isolated} is to be deleted and the right-hand side node is to be created. Hence, by the dangling condition, the rule is applicable only to a red root node. If \texttt{isolated} is not applicable, the node rooted by \texttt{mark} is not isolated and the loop is terminated by the \texttt{break} command. Finally, the rule \texttt{root} checks if a red root exists in the host graph, which is the case if and only if the application of \texttt{isolated} failed.

\definecolor{gp2pink}{RGB}{255, 153, 238}

\begin{figure}[!ht]
    \begin{mdframed}[linewidth=1pt]
    \begin{verbatim}
Main = (mark; try isolated else break)!; if root then fail
    \end{verbatim}
    
    \begin{tikzpicture}
    \tikzstyle{every node}=[font=\ttfamily]
    
    \draw (0.9,0.75) node[align=left] {mark(x:list)};
    \node[rectangle, thick, thick, rounded corners=7, draw=black, minimum size=5mm, label=below:\tiny\tiny 1](a2) at (0,0){x};
    \draw (1,0) node[] {$\Rightarrow$};
    \node[rectangle, thick, rounded corners=7, draw=black, minimum size=5mm,  double, double distance=1pt, fill=red!60, label=below:\tiny\tiny1](a2) at (2,0){x};
    \draw[->, thick];
    \draw (3.2,1) -- (3.2,-0.5);
    \end{tikzpicture}
    \hspace{1em}
    \begin{tikzpicture}
    \tikzstyle{every node}=[font=\ttfamily]
    \draw (1.3,0.75) node[align=left] {isolated(x:list)};
    \node[rectangle, thick, rounded corners=7, draw=black, minimum size=5mm,  double, double distance=1pt, fill=red!60, label=below:\tiny\tiny](a2) at (0,0){x};
    \draw (1,0) node[] {$\Rightarrow$};
    \node[rectangle, thick, rounded corners=7, draw=black, minimum size=5mm, fill=red!60, label=below:\tiny\tiny](a2) at (2,0){x};
    \draw (3.4,1) -- (3.4,-0.5);
    \draw (0,-0.62) -- (0,-0.62);
    \draw[->, thick];
    \end{tikzpicture}
    \hspace{1em}
    \begin{tikzpicture}
    \tikzstyle{every node}=[font=\ttfamily]
     \draw (0.9,0.75) node[align=left] {root(x:list)};
    \node[rectangle, thick, thick, rounded corners=7, draw=black, minimum size=5mm, fill=red!60,  double, double distance=1pt,  label=below:\tiny\tiny 1](a2) at (0,0){x};
    \draw (1,0) node[] {$\Rightarrow$};
    \node[rectangle, thick, rounded corners=7, draw=black, minimum size=5mm, double, double distance=1pt, fill=red!60, label=below:\tiny\tiny1](a2) at (2,0){x};
    \draw[->, thick];
    \end{tikzpicture}
    \end{mdframed}
    \caption{The program \texttt{is-discrete}.}
    \label{fig:is-discrete}
    \end{figure}

\begin{figure}[!ht]
    \centering
    \begin{tikzpicture}
    \begin{axis}[
      xlabel=number of nodes,
      ylabel=runtime (ms), ylabel style={above=0.2mm},
      width=9.2cm,height=7.2cm,
      legend style={at={(1.525,0.5)}},
      ymajorgrids=true,
      grid style=dashed]
      \addplot[color=plot8, mark=square*] table [y=time, x=n]{Figures/Benchmarks/is-dis.dat};
      \addlegendentry{2020 compiler}
      \addplot[color=plot7, mark=square*] table [y=time, x=n]{Figures/Benchmarks/is-dis-new.dat};
      \addlegendentry{New compiler}
    \end{axis}  
    \end{tikzpicture}
    \caption{Measured performance of \texttt{is-discrete} on discrete graphs under the 2020 compiler and the new compiler.}
    \label{fig:bench-is-dis}
    \end{figure}
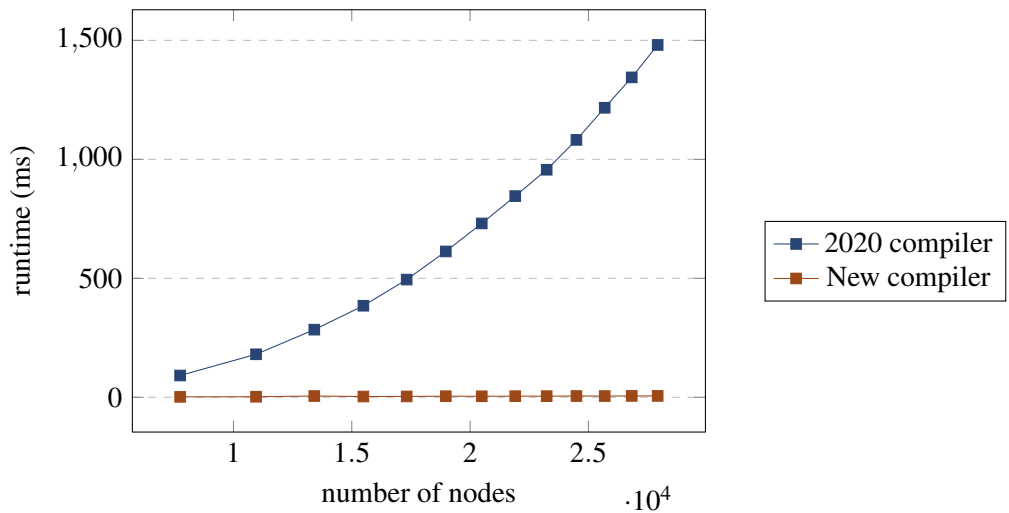

The 2020 compiler matched the rule \texttt{mark} with a complexity of $\mathrm{O}(n)$, where $n$ is the number of nodes in the host graph. In order to find an unmarked node, the matching algorithm had to iterate through the linked list containing all nodes. As the loop body is executed at most $n$ times, the overall complexity of \texttt{is-concrete} was $\mathrm{O}(n^2)$, as illustrated by the timing diagram in Figure \ref{fig:bench-is-dis}.

\section{Second Compiler Enhancement}

To overcome the problem described in Section \ref{ss:motiv}, the new graph data structure stores host-graph nodes in five global linked lists. Each list corresponds to one of the node marks red, green, blue, grey or unmarked, and holds all nodes with that mark. The first element of each linked list can be accessed in constant time and hence, for example, the rule \texttt{mark} in Figure \ref{fig:is-discrete} can be matched in constant time. The matching algorithm inspects the first element in the linked list of unmarked nodes. As the left-hand side of \texttt{mark} requires an unmarked node with an arbitrary label, the first node in the list will match. In case the list of unmarked nodes is empty, the host graph does not contain any unmarked node and thus the application of \texttt{mark} fails.

As a result of these changes, the time complexity of the program \texttt{is-discrete} under the new compiler is reduced to $\mathrm{O}(n)$. Figure \ref{fig:bench-is-dis} highlights the difference in runtimes of the program \texttt{is-discrete} run under both compilers.

\begin{figure}
\begin{center}
\begin{tabular}{l|c}
              
  unmarked                  & \dots \\
  \hline
  \textcolor{gp2grey}{grey} & \dots \\
  \hline
  \red{red}                 & \dots \\
  \hline
  \green{green}             & \dots \\
  \hline
  \blue{blue}               & \dots 
  
\end{tabular}  
    \caption{Array of linked lists of nodes.}
    \label{fig:node-lists}
\end{center}
\end{figure} 

\definecolor{modified}{gray}{0.9}
\definecolor{modified2}{rgb}{0.88, 1, 1}
\definecolor{added}{rgb}{0.88,1,1}

\begin{figure}[!t]
\begin{center}
\scalebox{.85}{
\begin{tabular}{l|l|l}
Procedure               & Description                                                    & Complexity  \\ \hline\hline
\texttt{alreadyMatched}    & Test if the given item has been matched in the host graph.     & \(O(1)\)    \\
\texttt{clearMatched}      & Clear the \texttt{is matched} flag for a given item.              & \(O(1)\)    \\
\texttt{setMatched}        & Set the \texttt{is matched} flag for a given item.                & \(O(1)\)    \\
\rowcolor{modified}\texttt{firstHostNode(m)}     & \textbf{Fetch the first node of mark \texttt{m} in the host graph.}                        & \(O(1)\)    \\
\rowcolor{modified}\texttt{nextHostNode(m)}      & \textbf{Given a node of mark \texttt{m}, fetch the next node of mark \texttt{m} in the host graph.}           & \(O(1)\)    \\
\texttt{firstHostRootNode\,\,\,} & Fetch the first root node in the host graph.                   & \(O(1)\)    \\
\texttt{nextHostRootNode\,\,\,}  & Given a root node, fetch the next root node in the host graph. & \(O(1)\)    \\
\rowcolor{modified2}\texttt{firstInEdge(m)}       & \textbf{Given a node, fetch the first incoming edge of mark \texttt{m}}.                 & \(O(1)\)    \\
\rowcolor{modified2}\texttt{nextInEdge(m)}        & \textbf{Given a node and an edge of mark \texttt{m}, fetch the next incoming edge of mark \texttt{m}}.\,\,\,         & \(O(1)\)    \\
\rowcolor{modified2}\texttt{firstOutEdge(m)}      & \textbf{Given a node, fetch the first outgoing edge of mark \texttt{m}}.                   & \(O(1)\)    \\
\rowcolor{modified2}\texttt{nextOutEdge(m)}       & \textbf{Given a node and an edge of mark \texttt{m}, fetch the next outgoing edge of mark \texttt{m}}.        & \(O(1)\)    \\
\rowcolor{modified2}\texttt{firstLoop(m)}       & \textbf{Given a node, fetch the first loop edge of mark \texttt{m}}.        & \(O(1)\)    \\
\rowcolor{modified2}\texttt{nextLoop(m)}       & \textbf{Given a node and an edge of mark \texttt{m}, fetch the next loop edge of mark \texttt{m}}.        & \(O(1)\)    \\
\texttt{getInDegree}       & Given a node, fetch its incoming degree.                       & \(O(1)\)    \\
\texttt{getOutDegree}      & Given a node, fetch its outgoing degree.                       & \(O(1)\)    \\
\texttt{getMark}           & Given a node or edge, fetch its mark.                          & \(O(1)\)    \\
\texttt{isRooted}          & Given a node, determine if it is rooted.                       & \(O(1)\)    \\
\texttt{getSource}         & Given an edge, fetch the source node.                          & \(O(1)\)    \\
\texttt{getTarget}         & Given an edge, fetch the target node.                          & \(O(1)\)    \\
\texttt{parseInputGraph}   & Parse and load the input graph into memory: the host graph.    & \(O(n)\)    \\
\texttt{printHostGraph}    & Write the current host graph state as output.                  & \(O(n)\)    \\
\end{tabular}
}
\end{center}
\caption{Updated runtime complexity assumptions. Procedures modified in this paper are highlighted in grey. Procedures modified in \cite{ismaili2024linear} are highlighted in light blue. $n$ is the size of the input.}
\label{fig:complexity-assumptions}
\end{figure}

To reason about programs, we need to make assumptions on the complexity of certain elementary operations. Figure \ref{fig:complexity-assumptions} shows the complexity of basic procedures of the C code generated by the GP\,2 compiler, adapted from \cite{campbell2022fast}. The grey rows indicate existing procedures updated by the changes introduced in this paper. The time complexities are consistent with the runtimes observed in all our case studies with the new compiler. 

\label{s:ce1}

\section{Case Study: Recognising Acyclic Graphs}


Checking whether a given graph contains a directed cycle is a basic problem in the area of graph algorithms \cite{Skiena20a}. A \gp{} program solving this problem is given in \cite{campbell2022fast}, but to run in linear time it requires input graphs of bounded node degree. The same paper contains a program for the related problem of recognising binary DAGs, which are acyclic graphs in which each node has at most two outgoing edges. This program has a linear runtime on arbitrary input graphs but is destructive in that the input graph is partially or totally deleted. 


\subsection{Program}

The program \texttt{is-dag} in Figure \ref{fig:is-dag-fig} recognises acylic graphs with respect to the following specification.

\begin{description}
\item[\textbf{Input:}] An arbitrary \gp{} host graph such that
\begin{enumerate}
        \item each node is non-rooted and marked grey, and
        \item each edge is unmarked.
    \end{enumerate}
\item[\textbf{Output:}] If the input graph is acyclic, a host graph that is isomorphic to the input graph up to marks. Otherwise \emph{failure}.
\end{description}

Strictly speaking, this program is destructive in case the input graph is cyclic because the \texttt{fail} command  in the procedure \texttt{Check}\
does not return an output graph. However, \texttt{is-dag} could easily be made completely non-destructive by replacing the \texttt{fail} command with a rule creating a distinct structure (such as an unmarked node) which signals the existence of a cycle. 

\definecolor{gp2pink}{RGB}{255, 153, 238}

\begin{figure}[!ht]
    \begin{mdframed}[linewidth=1pt]
    \begin{verbatim}
Main  = (init; DFS!; try unroot else break)!; Check
DFS   = try next_edge then (try {move, ignore} else (set_flag; break)) 
        else (try loop; try back else break)
Check = if flag then fail
    \end{verbatim}
    
    \begin{tikzpicture}
    \tikzstyle{every node}=[font=\ttfamily]
    
    \draw (0.9,0.75) node[align=left] {init(x:list)};
    \node[rectangle, thick, thick, rounded corners=7, draw=black, minimum size=5mm, fill=gray!50, label=below:\tiny\tiny 1](a2) at (0,0){x};
    \draw (1,0) node[] {$\Rightarrow$};
    \node[rectangle, thick, rounded corners=7, draw=black, minimum size=5mm,  double, double distance=1pt, fill=red!60, label=below:\tiny\tiny1](a2) at (2,0){x};
    \draw[->, thick];
    \draw (3,1) -- (3,-0.5);
    \end{tikzpicture}
    \hspace{1em}
    \begin{tikzpicture}
    \tikzstyle{every node}=[font=\ttfamily]
    \draw (1.1,0.75) node[align=left] {unroot(x:list)};
    \node[rectangle, thick, rounded corners=7, draw=black, minimum size=5mm,  double, double distance=1pt, fill=red!60, label=below:\tiny\tiny1](a2) at (0,0){x};
    \draw (1,0) node[] {$\Rightarrow$};
    \node[rectangle, thick, rounded corners=7, draw=black, minimum size=5mm, fill=cyan, label=below:\tiny\tiny 1](a2) at (2,0){x};
    \draw[->, thick];
    \draw (3,1) -- (3,-0.5);
    \end{tikzpicture}
    \hspace{1em}
    \begin{tikzpicture}
    \tikzstyle{every node}=[font=\ttfamily]
    \draw (1.3,0.75) node[align=left] {set\_flag(x:list)};
    \node[rectangle, thick, rounded corners=7, draw=black, minimum size=5mm,  double, double distance=1pt, fill=red!60, label=below:\tiny\tiny1](a2) at (0,0){x};
    \draw (1,0) node[] {$\Rightarrow$};
    \node[rectangle, thick, rounded corners=7, draw=black, minimum size=5mm, double, double distance=1pt, fill=green!60, label=below:\tiny\tiny 1](a2) at (2,0){x};
    \draw[->, thick];
    \draw (3.3,1) -- (3.3,-0.5);
    \end{tikzpicture}
    \hspace{0.2em}
    \begin{tikzpicture}
    \tikzstyle{every node}=[font=\ttfamily]
    \draw (0.9,0.75) node[align=left] {flag(x:list)};
    \node[rectangle, thick, rounded corners=7, draw=black, minimum size=5mm, double, double distance=1pt, fill=green!60, label=below:\tiny\tiny1](a2) at (0,0){x};
    \draw (1,0) node[] {$\Rightarrow$};
    \node[rectangle, thick, rounded corners=7, draw=black, minimum size=5mm, double, double distance=1pt, fill=green!60, label=below:\tiny\tiny 1](a2) at (2,0){x};
    \draw[->, thick];
    \end{tikzpicture}
    
    \begin{tikzpicture}
    \tikzstyle{every node}=[font=\ttfamily]
    \draw (1.8,0.75) node[align=left] {next\_edge(x,y,z:list)};
    \node[rectangle, thick, rounded corners=7, draw=black, minimum size=5mm,  double, double distance=1pt, fill=red!60, label=below:\tiny\tiny1](a1) at (0,0){x};
    \node[rectangle, thick, rounded corners=7, draw=black, minimum size=5mm, fill=gp2pink, label=below:\tiny\tiny2](a2) at (2,0){y};
    \draw (3,0) node[] {$\Rightarrow$};
    \node[rectangle, thick, rounded corners=7, draw=black, minimum size=5mm, double, double distance=1pt, fill=red!60, label=below:\tiny\tiny1](a3) at (4,0){x};
    \node[rectangle, thick, rounded corners=7, draw=black, minimum size=5mm, fill=gp2pink, label=below:\tiny\tiny2](a4) at (6,0){y};
    \draw[->, line width=1.2pt] (a1) edge[black] node[above, color = black]{z} (a2) (a3) edge[red] node[above, color = black]{z} (a4);
    \draw (7,1) -- (7,-0.5);
    \end{tikzpicture}
    \hspace{1em}
    \begin{tikzpicture}
    \tikzstyle{every node}=[font=\ttfamily]
    \draw (1.45,0.75) node[align=left] {ignore(x,y,z:list)};
    \node[rectangle, thick, rounded corners=7, draw=black, minimum size=5mm,  double, double distance=1pt, fill=red!60, label=below:\tiny\tiny1](a1) at (0,0){x};
    \node[rectangle, thick, rounded corners=7, draw=black, minimum size=5mm, fill=cyan, label=below:\tiny\tiny2](a2) at (2,0){y};
    \draw (3,0) node[] {$\Rightarrow$};
    \node[rectangle, thick, rounded corners=7, draw=black, minimum size=5mm, double, double distance=1pt, fill=red!60, label=below:\tiny\tiny1](a3) at (4,0){x};
    \node[rectangle, thick, rounded corners=7, draw=black, minimum size=5mm, fill=cyan, label=below:\tiny\tiny2](a4) at (6,0){y};
    \draw[->, line width=1.2pt] (a1) edge[red] node[above, color = black]{z} (a2) (a3) edge[cyan] node[above, color = black]{z} (a4);
    \end{tikzpicture}
    
    \begin{tikzpicture}
    \tikzstyle{every node}=[font=\ttfamily]
    \draw (1.3,0.75) node[align=left] {move(x,y,z:list)};
    \node[rectangle, thick, rounded corners=7, draw=black, minimum size=5mm,  double, double distance=1pt, fill=red!60, label=below:\tiny\tiny1](a1) at (0,0){x};
    \node[rectangle, thick, rounded corners=7, draw=black, minimum size=5mm, fill=gray!50, label=below:\tiny\tiny2](a2) at (2,0){y};
    \draw (3,0) node[] {$\Rightarrow$};
    \node[rectangle, thick, rounded corners=7, draw=black, minimum size=5mm, fill=red!60, label=below:\tiny\tiny1](a3) at (4,0){x};
    \node[rectangle, thick, rounded corners=7, draw=black, minimum size=5mm, double, double distance=1pt, fill=red!60, label=below:\tiny\tiny2](a4) at (6,0){y};
    \draw[->, line width=1.2pt] (a1) edge[red] node[above, color = black]{z} (a2) (a3) edge[dashed] node[above, color = black]{z} (a4);
    \draw (7,1) -- (7,-0.5);
    \end{tikzpicture}
    \hspace{1em}
    \begin{tikzpicture}
    \tikzstyle{every node}=[font=\ttfamily]
    \draw (1.3,0.75) node[align=left] {back(x,y,z:list)};
    \node[rectangle, thick, rounded corners=7, draw=black, minimum size=5mm, fill=red!60, label=below:\tiny\tiny1](a1) at (0,0){x};
    \node[rectangle, thick, rounded corners=7, draw=black, minimum size=5mm, double, double distance=1pt, fill=red!60, label=below:\tiny\tiny2](a2) at (2,0){y};
    \draw (3,0) node[] {$\Rightarrow$};
    \node[rectangle, thick, rounded corners=7, draw=black, minimum size=5mm, double, double distance=1pt, fill=red!60, label=below:\tiny\tiny1](a3) at (4,0){x};
    \node[rectangle, thick, rounded corners=7, draw=black, minimum size=5mm, fill=cyan, label=below:\tiny\tiny2](a4) at (6,0){y};
    \draw[->, line width=1.2pt] (a1) edge[dashed] node[above, color = black]{z} (a2) (a3) edge[cyan] node[above, color = black]{z} (a4);
    \end{tikzpicture}

    \begin{tikzpicture}
    \tikzstyle{every node}=[font=\ttfamily]
    \draw (1.1,0.65) node[align=left] {loop(x,z:list)};
    \node[rectangle, thick, rounded corners=7, draw=black, minimum size=5mm,  double, double distance=1pt, fill=red!60, label=below:\tiny\tiny1](a2) at (0,0){x};
    \draw (1,0) node[] {$\Rightarrow$};
    \node[rectangle, thick, rounded corners=7, draw=black, minimum size=5mm, double, double distance=1pt, fill=green!60, label=below:\tiny\tiny1](a3) at (2,0){x};
    \draw[->, line width=1.2pt] (a2) to [out=330,in=300,looseness=8] node[right, color = black]{z} (a2);
    \draw[->, line width=1.2pt] (a3) to [out=330,in=300,looseness=8] node[right, color = black]{z} (a3);
    \end{tikzpicture}
    
    \end{mdframed}
    \caption{The program \texttt{is-dag}.}
    \label{fig:is-dag-fig}
    \end{figure}

Figure \ref{fig:is-dag-ex} illustrates an execution of \texttt{is-dag} on a cyclic input graph, and Figure \ref{fig:is-dag-ex-acyclic}, on an acylic graph. The program implements a directed DFS (depth-first search) of the host graph that marks the visited nodes red or blue, where the red nodes are currently being visited. When modelling the DFS with a stack, red nodes are currently on the stack while blue nodes have been previously on the stack and were popped because they require no further visits. Moreover, the top of the stack is a root node.

It is an invariant of \texttt{is-dag} that there is at most one root in the host graph throughout the program's execution. The graph contains a cycle if and only if the search finds an edge from the root to a red node. In fact, if $u$ is the root and $v$ is a red node adjacent to $u$ via an edge from $u$ to $v$, then there must exist a directed path from $v$ to $u$. Hence a directed cycle has been found.

\begin{figure}
    \centering
    \usetikzlibrary{overlay-beamer-styles}
\newcommand{\mtt}{\mathtt}
\newcommand{\mrm}{\mathrm}

\definecolor{gp2green}{RGB}{153, 255, 170}
\definecolor{gp2blue}{RGB}{120, 161, 242}
\definecolor{gp2red}{RGB}{233, 73, 87}
\definecolor{gp2pink}{RGB}{255, 153, 238}
\definecolor{gp2grey}{RGB}{210, 210, 210}
\definecolor{plot1}{RGB}{70, 116, 193}
\definecolor{plot2}{RGB}{235, 125, 60}
\definecolor{plot3}{RGB}{165, 165, 165}
\definecolor{plot4}{RGB}{252, 190, 45}
\definecolor{plot5}{RGB}{94, 156, 210}
\definecolor{plot6}{RGB}{113, 171, 77}
\definecolor{plot7}{RGB}{156, 72, 25}
\definecolor{plot8}{RGB}{40, 69, 117}

\tikzset{gp2 node/.style={draw, circle, thick, minimum width=0.64cm}}
\tikzset{root node/.style={draw, circle, thick, minimum width=0.64cm, double, double distance=0.3mm}}

\scalebox{0.95}{%
\begin{tabular}{ccccccc}
\begin{minipage}{2.1cm}
\centering
\begin{tikzpicture}[scale=0.6]
	\node (a) at (-0.500,0.000)   [draw,circle,thick, fill=gray!50] {\,};
	\node (b) at (1.000,0.000)   [draw,circle,thick, fill=gray!50] {\,};
	\node (c) at (0.250,2.000)   [draw,circle,thick, fill=gray!50] {\,};
        \node (d) at (-1.000,2.000)   [draw,circle,thick, fill=gray!50] {\,};
	\draw (a) edge[->, thick] (b)
              (b) edge[->, thick] (c)
              (c) edge[->, thick] (a)
              (d) edge[->, thick] (a);
\end{tikzpicture}
\end{minipage}
&
$\Rightarrow_{\mtt{init}}$
&
\begin{minipage}{2.1cm}
\centering
\begin{tikzpicture}[scale=0.6]
	\node (a) at (-0.500,0.000)   [draw,circle,thick, fill=gray!50] {\,};
	\node (b) at (1.000,0.000)   [draw,circle,thick, fill=gray!50] {\,};
	\node (c) at (0.250,2.000)  [draw,circle,thick, fill=gray!50] {\,};
        \node (d) at (-1.000,2.000)   [root node,fill=red!60, inner sep=0pt, minimum size=0.4cm] {\,};
	\draw (a) edge[->, thick] (b)
              (b) edge[->, thick] (c)
              (c) edge[->, thick] (a)
              (d) edge[->, thick] (a);
\end{tikzpicture}
\end{minipage}
&
$\Rightarrow_{\mtt{nx\_dg}}$
&
\begin{minipage}{2.1cm}
\centering
\begin{tikzpicture}[scale=0.6]
	\node (a) at (-0.500,0.000)   [draw,circle,thick, fill=gray!50] {\,};
	\node (b) at (1.000,0.000)   [draw,circle,thick, fill=gray!50] {\,};
	\node (c) at (0.250,2.000)  [draw,circle,thick, fill=gray!50] {\,};
        \node (d) at (-1.000,2.000)   [root node,fill=red!60, inner sep=0pt, minimum size=0.4cm] {\,};
	\draw (a) edge[->, thick] (b)
              (b) edge[->, thick] (c)
              (c) edge[->, thick] (a)
              (d) edge[->, thick, red] (a);
\end{tikzpicture}
\end{minipage}
&
$\Rightarrow_{\mtt{mv}}$
&
\begin{minipage}{2.1cm}
\centering
\begin{tikzpicture}[scale=0.6]
	\node (a) at (-0.500,0.000)  [root node,fill=red!60, inner sep=0pt, minimum size=0.4cm] {\,};
	\node (b) at (1.000,0.000)   [draw,circle,thick, fill=gray!50] {\,};
	\node (c) at (0.250,2.000)  [draw,circle,thick, fill=gray!50] {\,};
        \node (d) at (-1.000,2.000)  [draw,circle,thick, fill=red!60] {\,};
	\draw (a) edge[->, thick] (b)
              (b) edge[->, thick] (c)
              (c) edge[->, thick] (a)
              (d) edge[->, thick, dashed] (a);
\end{tikzpicture}
\end{minipage}
\\
\\
 &&&&&& $\Downarrow_{\mtt{nx\_dg}}$
\\[1ex]
\begin{minipage}{2.1cm}
\centering
\begin{tikzpicture}[scale=0.6]
	\node (a) at (-0.500,0.000){};  	
    \node (a) at (-0.500,0.000)  [draw,circle,thick, fill=red!60] {\,};
	\node (b) at (1.000,0.000)   [draw,circle,thick, fill=red!60] {\,};
	\node (c) at (0.250,2.000)  [root node,fill=red!60, inner sep=0pt, minimum size=0.4cm] {\,};
        \node (d) at (-1.000,2.000)  [draw,circle,thick, fill=red!60] {\,};
	\draw (a) edge[->, thick, dashed] (b)
              (b) edge[->, thick, dashed] (c)
              (c) edge[->, thick] (a)
              (d) edge[->, thick, dashed] (a);
\end{tikzpicture}
\end{minipage}
&
$\Leftarrow_{\mtt{mv}}$
&
\begin{minipage}{2.1cm}
\centering
\begin{tikzpicture}[scale=0.6]
	\node (a) at (-0.500,0.000)  [draw,circle,thick, fill=red!60] {\,};
	\node (b) at (1.000,0.000) [root node,fill=red!60, inner sep=0pt, minimum size=0.4cm] {\,};
	\node (c) at (0.250,2.000)  [draw,circle,thick, fill=gray!50] {\,};
        \node (d) at (-1.000,2.000)  [draw,circle,thick, fill=red!60] {\,};
	\draw (a) edge[->, thick, dashed] (b)
              (b) edge[->, thick, red] (c)
              (c) edge[->, thick] (a)
              (d) edge[->, thick, dashed] (a);
\end{tikzpicture}
\end{minipage}
&
$\Leftarrow_{\mtt{nx\_dg}}$
&
\begin{minipage}{2.1cm}
\centering
\begin{tikzpicture}[scale=0.6]
	\node (a) at (-0.500,0.000)  [draw,circle,thick, fill=red!60] {\,};
	\node (b) at (1.000,0.000)   [root node,fill=red!60, inner sep=0pt, minimum size=0.4cm] {\,};
	\node (c) at (0.250,2.000)  [draw,circle,thick, fill=gray!50] {\,};
        \node (d) at (-1.000,2.000)  [draw,circle,thick, fill=red!60] {\,};
	\draw (a) edge[->, thick, dashed] (b)
              (b) edge[->, thick] (c)
              (c) edge[->, thick] (a)
              (d) edge[->, thick, dashed] (a);
\end{tikzpicture}
\end{minipage}
&
$\Leftarrow_{\mtt{mv}}$
&
\begin{minipage}{2.1cm}
\centering
\begin{tikzpicture}[scale=0.6]
	\node (a) at (-0.500,0.000)  [root node,fill=red!60, inner sep=0pt, minimum size=0.4cm] {\,};
	\node (b) at (1.000,0.000)   [draw,circle,thick, fill=gray!50] {\,};
	\node (c) at (0.250,2.000)  [draw,circle,thick, fill=gray!50] {\,};
        \node (d) at (-1.000,2.000)  [draw,circle,thick, fill=red!60] {\,};
	\draw (a) edge[->, thick, red] (b)
              (b) edge[->, thick] (c)
              (c) edge[->, thick] (a)
              (d) edge[->, thick, dashed] (a);
\end{tikzpicture}
\end{minipage}
\\
\\
$\Downarrow_{\mtt{nx\_dg}}$ &&&&&&
\\[1ex]
\begin{minipage}{2.1cm}
\centering
\begin{tikzpicture}[scale=0.6]
	\node (a) at (-0.500,0.000)  [draw,circle,thick, fill=red!60] {\,};
	\node (b) at (1.000,0.000)   [draw,circle,thick, fill=red!60] {\,};
	\node (c) at (0.250,2.000)  [root node,fill=red!60, inner sep=0pt, minimum size=0.4cm] {\,};
        \node (d) at (-1.000,2.000)  [draw,circle,thick, fill=red!60] {\,};
	\draw (a) edge[->, thick, dashed] (b)
              (b) edge[->, thick, dashed] (c)
              (c) edge[->, thick, red] (a)
              (d) edge[->, thick, dashed] (a);
\end{tikzpicture}
\end{minipage}
&
$\Rightarrow_{\mtt{set\_flag}}$
&
\begin{minipage}{2.1cm}
\centering
\begin{tikzpicture}[scale=0.6]
	\node (a) at (-0.500,0.000)  [draw,circle,thick, fill=red!60] {\,};
	\node (b) at (1.000,0.000)   [draw,circle,thick, fill=red!60] {\,};
	\node (c) at (0.250,2.000)  [root node,fill=green!60, inner sep=0pt, minimum size=0.4cm] {\,};
        \node (d) at (-1.000,2.000)  [draw,circle,thick, fill=red!60] {\,};
	\draw (a) edge[->, thick, dashed] (b)
              (b) edge[->, thick, dashed] (c)
              (c) edge[->, thick, red] (a)
              (d) edge[->, thick, dashed] (a);
\end{tikzpicture}
\end{minipage}
&
$\Rightarrow_{\mtt{flag}}$
&
\begin{minipage}{2.1cm}
\centering
\begin{tikzpicture}[scale=0.6]
	\node (a) at (-0.500,0.000)  [draw,circle,thick, fill=red!60] {\,};
	\node (b) at (1.000,0.000)   [draw,circle,thick, fill=red!60] {\,};
	\node (c) at (0.250,2.000)  [root node,fill=green!60, inner sep=0pt, minimum size=0.4cm] {\,};
        \node (d) at (-1.000,2.000)  [draw,circle,thick, fill=red!60] {\,};
	\draw (a) edge[->, thick, dashed] (b)
              (b) edge[->, thick, dashed] (c)
              (c) edge[->, thick, red] (a)
              (d) edge[->, thick, dashed] (a);
\end{tikzpicture}
\end{minipage}
&
$\Rightarrow_{\mtt{fail}}$
&
\begin{minipage}{2.1cm}
\centering
\begin{tikzpicture}[scale=0.6]
\tikzstyle{every node}=[]

\node[](v1) at (0.5,0.6){Failure};
\node[](v2) at (0,0){};
\end{tikzpicture}
\end{minipage}
\end{tabular}}
    \caption{Sample execution of \texttt{is-dag} on a cyclic graph.}
    \label{fig:is-dag-ex}
\end{figure}
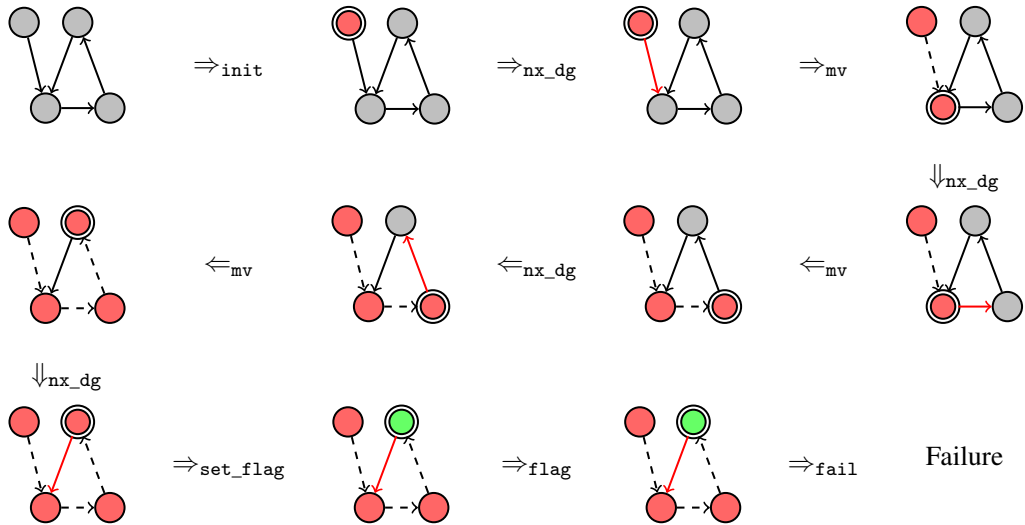

\begin{figure}
    \centering
    \input{Figures/Diagrams/is-dag-ex-acyclic}
    \caption{Sample execution of \texttt{is-dag} on an acyclic graph.}
    \label{fig:is-dag-ex-acyclic}
\end{figure}

Consider the loop \texttt{(init; DFS!; try unroot else break)!} of \texttt{is-dag}'s main procedure. Rule \texttt{init} selects an arbitrary grey node as a root to start a directed DFS. The loop \texttt{DFS!} moves the root in depth-first fashion through the host graph. The procedure uses a \texttt{try-else} command to find any unprocessed (that is, unmarked) edge outgoing from the root. It does this by calling \texttt{next\_edge}.\footnote{In the programs of this paper, we use the magenta colour to represent the wildcard mark \texttt{any}.} If there is such an edge, the rule marks it red so that it can be uniquely identified by the rest of the procedure. If no such edge exits, the root can no longer move forward and the \texttt{else} statement is invoked instead. 

After a successful application of \texttt{next\_edge}, the root is adjacent to either (1) a grey node, (2) a blue node, or (3) a red node. In case (1), the rule \texttt{move} moves the root to the grey node, marks it red and dashes the traversed edge. Dashed edges represent the path followed by the directed DFS. In case (2), the red edge is marked blue by the rule \texttt{ignore} so that it can no longer be matched by \texttt{next\_edge}. In case (3), neither \texttt{move} nor \texttt{ignore} is applicable so that \texttt{set\_flag} marks the root green, indicating the existence of a cycle. 

If \texttt{next\_edge} is not applicable, the command sequence \texttt{(try loop; try back else break)} is executed. Rule \texttt{loop} checks whether there is a loop attached to the root. If this is the case, the rule marks the root green, similar to \texttt{set\_flag}. Then rule \texttt{back} is tried which implements the \textit{pop} operation in the above mentioned stack model. The rule moves the root backwards along an incoming dashed edge. If no incoming dashed edge is present, the root must be the only element on the stack so that the \texttt{break} command terminates the loop \texttt{DFS!}. 

Upon termination of \texttt{DFS!}, the rule \texttt{unroot} attempts to turn the root into an unrooted blue node. If this is not possible, the root must have been marked green by \texttt{set\_flag} or \texttt{loop}. This implies the existence of a cycle and hence the outer loop of \texttt{is-dag} is terminated. 

If rule \texttt{unroot} could be applied, there may still be nodes that have not been visited by the DFS. These are nodes that are not directly reachable from the initial nodes chosen so far. In this case the execution of the outer loop is continued until \texttt{init} is no longer applicable or  \texttt{unroot} fails. 

Finally, the procedure \texttt{Check} tests whether a green flag exists in the host graph. Should this be the case, a cycle was found and the command \texttt{fail} terminates the program with failure. Otherwise, the program terminates by returning a host graph which is isomorphic to the input graph up to the generated marks.

\subsection{Time Complexity}

All rules of the program \texttt{is-dag} apply in constant time under the complexity assumptions of the modified \gp{} compiler (Figure \ref{fig:complexity-assumptions}). The rule \texttt{init} applies in constant time since any grey-marked node is a match for the rule. As the generated graph data structure keeps a linked list of nodes for each node mark, the matching algorithm can select a grey node in constant time regardless of the number of non-grey nodes in the host graph.

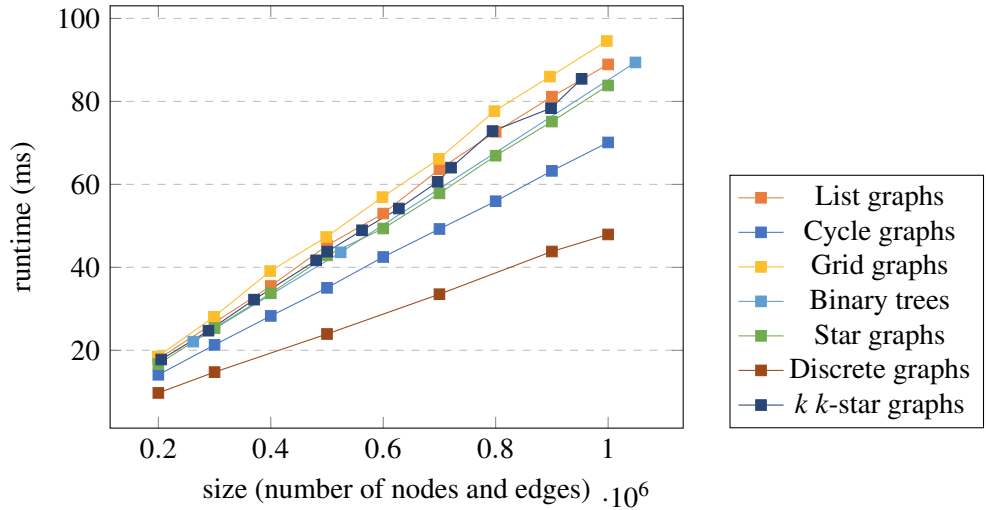
\begin{figure}[!ht]
    \centering
    \begin{tikzpicture}
    \begin{axis}[
      xlabel=size (number of nodes and edges),
      ylabel=runtime (ms), ylabel style={above=0.2mm},
      width=9.2cm,height=7.2cm,
      legend style={at={(1.525,0.6)}},
      ymajorgrids=true,
      grid style=dashed]
      \addplot[color=plot2, mark=square*] table [y=time, x=n]{Figures/Benchmarks/DAG/is-dag-list.dat};
      \addlegendentry{List graphs}
      \addplot[color=plot1, mark=square*] table [y=time, x=n]{Figures/Benchmarks/DAG/is-dag-cycle.dat};
      \addlegendentry{Cycle graphs}
      \addplot[color=plot4, mark=square*] table [y=time, x=n]{Figures/Benchmarks/DAG/is-dag-grid.dat};
      \addlegendentry{Grid graphs}
      \addplot[color=plot5, mark=square*] table [y=time, x=n]{Figures/Benchmarks/DAG/is-dag-tree.dat};
      \addlegendentry{Binary trees}
      \addplot[color=plot6, mark=square*] table [y=time, x=n]{Figures/Benchmarks/DAG/is-dag-star.dat};
      \addlegendentry{Star graphs}
      \addplot[color=plot7, mark=square*] table [y=time, x=n]{Figures/Benchmarks/DAG/is-dag-discrete.dat};
      \addlegendentry{Discrete graphs}
      \addplot[color=plot8, mark=square*] table [y=time, x=n]{Figures/Benchmarks/DAG/is-dag-k-star.dat};
      \addlegendentry{$k$ $k$-star graphs}
    \end{axis}  
    \end{tikzpicture}
    \caption{Measured performance of the program \texttt{is-dag} under the modified compiler.}
    \label{fig:bench-is-dag}
\end{figure}

It can be observed that nodes and edges are never remarked by a mark they previously had. Since all rules, except \texttt{flag} called at most once in \texttt{Check}, remark at least one element, the overall program runtime is linear in the size of the graph, i.e. $\mathrm{O}(n)$ where $n$ is the number of nodes and edges in the input graph. To support this claim, we conducted runtime experiments on various classes of bounded-degree graphs (Figures \ref{fig:graph-class-1b}, \ref{fig:graph-class-1c}, \ref{fig:graph-class-2b}, \ref{fig:graph-class-3b} and \ref{fig:graph-class-6bd}),  and unbounded-degree graphs (Figures \ref{fig:graph-class-2a} and \ref{fig:graph-class-6b}). The timing results are shown in Figure \ref{fig:bench-is-dag}.
\newline

\begin{center}\begin{center}
    \resizebox{0.82\linewidth}{!}{\centering
\begin{minipage}{3.93cm}
\centering
\begin{tikzpicture}[scale=0.7]
	\node (a) at (-1.500,1.333)  [draw,circle,thick,fill=gray!50] {\,};
	\node (b) at (0.000,1.333)   [draw,circle,thick,fill=gray!50] {\,};
	\node (c) at (1.500,1.333)   [draw,circle,thick,fill=gray!50] {\,};
	\node (d) at (-1.500,0.000)  [draw,circle,thick,fill=gray!50] {\,};
	\node (e) at (0.000,0.000)   [draw,circle,thick,fill=gray!50] {\,};
	\node (f) at (1.500,0.000)   [draw,circle,thick,fill=gray!50] {\,};
	\node (g) at (-1.500,-1.333) [draw,circle,thick,fill=gray!50] {\,};
	\node (h) at (0.000,-1.333)  [draw,circle,thick,fill=gray!50] {\,};
	\node (i) at (1.500,-1.333)  [draw,circle,thick,fill=gray!50] {\,};
	
	\draw (a) edge[->, thick] (b)
	      (a) edge[->, thick] (d)
	      (b) edge[->, thick] (c)
	      (b) edge[->, thick] (e)
	      (c) edge[->, thick] (f)
	      (d) edge[->, thick] (e)
	      (d) edge[->, thick] (g)
	      (e) edge[->, thick] (f)
	      (e) edge[->, thick] (h)
	      (f) edge[->, thick] (i)
	      (g) edge[->, thick] (h)
	      (h) edge[->, thick] (i);
\end{tikzpicture}
\captionof{figure}{Grid graph.}
\vspace{0.5cm}
\label{fig:graph-class-1b}
\end{minipage}
\begin{minipage}{3.93cm}
\centering
\begin{tikzpicture}[scale=0.7]
	\node (a) at (0.000,1.333)   [draw,circle,thick,fill=gray!50] {\,};
	\node (b) at (1.333,0.000)   [draw,circle,thick,fill=gray!50] {\,};
	\node (c) at (-1.333,0.000)  [draw,circle,thick,fill=gray!50] {\,};
	\node (d) at (2.000,-1.333)  [draw,circle,thick,fill=gray!50] {\,};
	\node (e) at (0.666,-1.333)  [draw,circle,thick,fill=gray!50] {\,};
	\node (f) at (-0.666,-1.333) [draw,circle,thick,fill=gray!50] {\,};
	\node (g) at (-2.000,-1.333) [draw,circle,thick,fill=gray!50] {\,};
	
	\draw (a) edge[->, thick] (b)
	      (a) edge[->, thick] (c)
	      (b) edge[->, thick] (d)
	      (b) edge[->, thick] (e)
	      (c) edge[->, thick] (f)
	      (c) edge[->, thick] (g);
\end{tikzpicture}
\captionof{figure}{Binary tree.}
\vspace{0.5cm}
\label{fig:graph-class-1c}
\end{minipage}
\begin{minipage}{3.93cm}
\centering
\begin{tikzpicture}[scale=0.7]
	\node (a) at (0.000,0.000)   [draw,circle,thick,fill=gray!50] {\,};
	\node (b) at (0.000,1.333)   [draw,circle,thick,fill=gray!50] {\,};
	\node (c) at (0.943,0.943)   [draw,circle,thick,fill=gray!50] {\,};
	\node (d) at (1.333,0.000)   [draw,circle,thick,fill=gray!50] {\,};
	\node (e) at (0.943,-0.943)  [draw,circle,thick,fill=gray!50] {\,};
	\node (f) at (0.000,-1.333)  [draw,circle,thick,fill=gray!50] {\,};
	\node (g) at (-0.943,-0.943) [draw,circle,thick,fill=gray!50] {\,};
	\node (h) at (-1.333,0.000)  [draw,circle,thick,fill=gray!50] {\,};
	\node (i) at (-0.943,0.943)  [draw,circle,thick,fill=gray!50] {\,};
	
	\draw (a) edge[->, thick] (b)
	      (c) edge[->, thick] (a)
	      (a) edge[->, thick] (d)
	      (e) edge[->, thick] (a)
	      (a) edge[->, thick] (f)
	      (g) edge[->, thick] (a)
	      (a) edge[->, thick] (h)
	      (i) edge[->, thick] (a);
\end{tikzpicture}
\captionof{figure}{Star graph.}
\vspace{0.5cm}
\label{fig:graph-class-2a}
\end{minipage}
\begin{minipage}{3.93cm}
\centering
\begin{tikzpicture}[scale=0.7]
    \node (a) at (0.0000,1.3333)   [draw,circle,thick,fill=gray!50] {\,};
	\node (b) at (1.1545,0.6666)   [draw,circle,thick,fill=gray!50] {\,};
	\node (c) at (1.1545,-0.6666)  [draw,circle,thick,fill=gray!50] {\,};
	\node (d) at (0.0000,-1.3334)  [draw,circle,thick,fill=gray!50] {\,};
	\node (e) at (-1.1545,-0.6666) [draw,circle,thick,fill=gray!50] {\,};
	\node (f) at (-1.1545,0.6666)  [draw,circle,thick,fill=gray!50] {\,};
	
	\draw (a) edge[->, thick] (b)
	      (b) edge[->, thick] (c)
	      (c) edge[->, thick] (d)
	      (d) edge[->, thick] (e)
	      (e) edge[->, thick] (f)
	      (f) edge[->, thick] (a);
\end{tikzpicture}

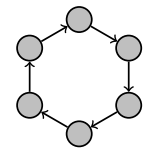
\captionof{figure}{Cycle graph.}
\vspace{0.5cm}
\label{fig:graph-class-2b}
\end{minipage}}
    \resizebox{1\linewidth}{!}{
\begin{minipage}{6.9cm}
    \centering
    \begin{tikzpicture}[scale=0.7]
        \node (a) at (-2.5,0)  [draw,circle,thick,fill=gray!50] {\,};
        \node (b) at (-0.8333,0)   [draw,circle,thick,fill=gray!50] {\,};
        \node (c) at (0.8333,0) [draw,circle,thick,fill=gray!50] {\,};
        \node (d) at (2.500,0)  [draw,circle,thick,fill=gray!50] {\,};
            \node (X) at (0,1.333){\,};
            \node (y) at (0,-1.333){\,};
        
        \draw (a) edge[->, thick] (b)
              (b) edge[->, thick] (c)
              (c) edge[->, thick] (d);
    \end{tikzpicture}
    
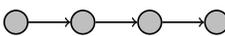
\captionof{figure}{Linked list.}
    \label{fig:graph-class-3b}
\end{minipage}
\begin{minipage}{6.9cm}
    \centering
    \begin{tikzpicture}[scale=0.7]
        \node (b) at (-0.8333,0)   [draw,circle,thick,fill=gray!50] {\,};
        \node (c) at (0.8333,0) [draw,circle,thick,fill=gray!50] {\,};
        \node (d) at (2.500,0)  [draw,circle,thick,fill=gray!50] {\,};
            \node (X) at (0,1.333){\,};
            \node (y) at (0,-1.333){\,};

    \end{tikzpicture}
    
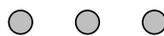
\captionof{figure}{Discrete graph.}
    \label{fig:graph-class-6bd}
\end{minipage}
\begin{minipage}{6.9cm}
    \centering
    \begin{tikzpicture}[scale=0.7]
	\node (a) at (0.000,0.000)   [draw,circle,thick,fill=gray!50] {\,};
	\node (c) at (0.943,0.943)   [draw,circle,thick,fill=gray!50] {\,};
	\node (f) at (0.000,-1.333)  [draw,circle,thick,fill=gray!50] {\,};
	\node (i) at (-0.943,0.943)  [draw,circle,thick,fill=gray!50] {\,};
	\draw 
	      (c) edge[->, thick] (a)
	      (f) edge[->, thick] (a)
	      (i) edge[->, thick] (a);

       \node (a2) at (3.000,0.000)   [draw,circle,thick,fill=gray!50] {\,};
	\node (c2) at (3.943,0.943)   [draw,circle,thick,fill=gray!50] {\,};
	\node (f2) at (3.000,-1.333)  [draw,circle,thick,fill=gray!50] {\,};
	\node (i2) at (2.057,0.943)  [draw,circle,thick,fill=gray!50] {\,};
	\draw 
	      (c2) edge[->, thick] (a2)
	      (f2) edge[->, thick] (a2)
	      (i2) edge[->, thick] (a2);

       \node (a3) at (6.000,0.000)   [draw,circle,thick,fill=gray!50] {\,};
	\node (c3) at (6.943,0.943)   [draw,circle,thick,fill=gray!50] {\,};
	\node (f3) at (6.000,-1.333)  [draw,circle,thick,fill=gray!50] {\,};
	\node (i3) at (5.057,0.943)  [draw,circle,thick,fill=gray!50] {\,};
	\draw 
	      (c3) edge[->, thick] (a3)
	      (f3) edge[->, thick] (a3)
	      (i3) edge[->, thick] (a3);
    \end{tikzpicture}
    
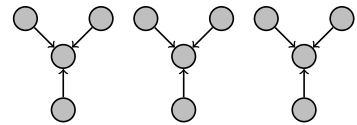
\captionof{figure}{$k$ $k$-star graphs.}
    \label{fig:graph-class-6b}
\end{minipage}}
    \end{center}
\end{center}

\section{Case Study: Numbering Connected Components}
A clear advantage of the new data structure is the ability to match an arbitrarily-labelled node of a particular mark (or determine that none exists) in constant time. A natural choice of a program that can be constructed under that paradigm is one that numbers all connected components of an input graph.

The program \texttt{component-numbering} from Figure \ref{fig:count-con-fig} appends a number to the list of each node of an input graph unique to the connected component the node belongs to with respect to the following specification.

\begin{description}
\item[\textbf{Input:}] An arbitrary \gp{} host graph such that
\begin{enumerate}
        \item each node is non-rooted and marked grey, and
        \item each edge is unmarked.
    \end{enumerate}
\item[\textbf{Output:}] A host graph structually isomorphic to the input graph where a number is appended to the list of each node, denoting the unique identifier of the connected component it belongs to.
\end{description}

\definecolor{gp2pink}{RGB}{255, 153, 238}

\begin{figure}[!ht]
    \begin{mdframed}[linewidth=1pt]
    \begin{verbatim}
Main = try init then DFS!; (try next then DFS! else break)!; try unroot
DFS  = (next_edge; {move, ignore})!; try back else break
    \end{verbatim}
    
    \begin{tikzpicture}
    \tikzstyle{every node}=[font=\ttfamily]
    
    \draw (0.9,0.75) node[align=left] {init(x:list)};
    \node[rectangle, thick, thick, rounded corners=7, draw=black, minimum size=5mm, fill=gray!50, label=below:\tiny\tiny 1](a2) at (0,0){x};
    \draw (1,0) node[] {$\Rightarrow$};
    \node[rectangle, thick, rounded corners=7, draw=black, minimum size=5mm,  double, double distance=1pt, fill=cyan, label=below:\tiny\tiny1](a2) at (2,0){x:1};
    \draw[-, thick];
    \draw (3,1) -- (3,-0.5);
    \end{tikzpicture}
    \hspace{1em}
    \begin{tikzpicture}
    \tikzstyle{every node}=[font=\ttfamily]
    \draw (1.1,0.75) node[align=left] {unroot(x:list)};
    \node[rectangle, thick, rounded corners=7, draw=black, minimum size=5mm,  double, double distance=1pt, fill=cyan, label=below:\tiny\tiny1](a2) at (0,0){x};
    \draw (1,0) node[] {$\Rightarrow$};
    \node[rectangle, thick, rounded corners=7, draw=black, minimum size=5mm, fill=cyan, label=below:\tiny\tiny 1](a2) at (2,0){x};
    \draw[-, thick];
    \draw (3,1) -- (3,-0.5);
    \end{tikzpicture}
    \hspace{1em}
    \begin{tikzpicture}
    \tikzstyle{every node}=[font=\ttfamily]
    \draw (1.65,0.75) node[align=left] {next(x,y:list,n:int)};
    \node[rectangle, thick, rounded corners=7, draw=black, minimum size=5mm, fill=cyan, double, distance=1pt, label=below:\tiny\tiny1](a1) at (0,0){x:n};
    \node[rectangle, thick, rounded corners=7, draw=black, minimum size=5mm, fill=gray!50, label=below:\tiny\tiny 1](a2) at (1,0){y};
    \draw (2.1,0) node[] {$\Rightarrow$};
    \node[rectangle, thick, rounded corners=7, draw=black, minimum size=5mm, fill=cyan, label=below:\tiny\tiny 1](a3) at (3.3,0){x:n};
    \node[rectangle, thick, rounded corners=7, draw=black, double, double distance=1pt, minimum size=5mm, fill=cyan, label=below:\tiny\tiny 1](a4) at (4.8,0){y:n+1};
    \draw[-, thick];
    \end{tikzpicture}
    
    \begin{tikzpicture}
    \tikzstyle{every node}=[font=\ttfamily]
    \draw (1.8,0.75) node[align=left] {next\_edge(x,y,z:list)};
    \node[rectangle, thick, rounded corners=7, draw=black, minimum size=5mm,  double, double distance=1pt, fill=cyan, label=below:\tiny\tiny1](a1) at (0,0){x};
    \node[rectangle, thick, rounded corners=7, draw=black, minimum size=5mm, fill=gp2pink, label=below:\tiny\tiny2](a2) at (2,0){y};
    \draw (3,0) node[] {$\Rightarrow$};
    \node[rectangle, thick, rounded corners=7, draw=black, minimum size=5mm, double, double distance=1pt, fill=cyan, label=below:\tiny\tiny1](a3) at (4,0){x};
    \node[rectangle, thick, rounded corners=7, draw=black, minimum size=5mm, fill=gp2pink, label=below:\tiny\tiny2](a4) at (6,0){y};
    \draw[-, line width=1.2pt] (a1) edge[black] node[above, color = black]{z} (a2) (a3) edge[red] node[above, color = black]{z} (a4);
    \draw (7,1) -- (7,-0.5);
    \end{tikzpicture}
    \hspace{1em}
    \begin{tikzpicture}
    \tikzstyle{every node}=[font=\ttfamily]
    \draw (1.45,0.75) node[align=left] {ignore(x,y,z:list)};
    \node[rectangle, thick, rounded corners=7, draw=black, minimum size=5mm,  double, double distance=1pt, fill=cyan, label=below:\tiny\tiny1](a1) at (0,0){x};
    \node[rectangle, thick, rounded corners=7, draw=black, minimum size=5mm, fill=cyan, label=below:\tiny\tiny2](a2) at (2,0){y};
    \draw (3,0) node[] {$\Rightarrow$};
    \node[rectangle, thick, rounded corners=7, draw=black, minimum size=5mm, double, double distance=1pt, fill=cyan, label=below:\tiny\tiny1](a3) at (4,0){x};
    \node[rectangle, thick, rounded corners=7, draw=black, minimum size=5mm, fill=cyan, label=below:\tiny\tiny2](a4) at (6,0){y};
    \draw[-, line width=1.2pt] (a1) edge[red] node[above, color = black]{z} (a2) (a3) edge[cyan] node[above, color = black]{z} (a4);
    \end{tikzpicture}
    
    \begin{tikzpicture}
    \tikzstyle{every node}=[font=\ttfamily]
    \draw (1.85,0.75) node[align=left] {move(x,y,z:list,i:int)};
    \node[rectangle, thick, rounded corners=7, draw=black, minimum size=5mm,  double, double distance=1pt, fill=cyan, label=below:\tiny\tiny1](a1) at (0,0){x:i};
    \node[rectangle, thick, rounded corners=7, draw=black, minimum size=5mm, fill=gray!50, label=below:\tiny\tiny2](a2) at (2,0){y};
    \draw (3,0) node[] {$\Rightarrow$};
    \node[rectangle, thick, rounded corners=7, draw=black, minimum size=5mm, fill=cyan, label=below:\tiny\tiny1](a3) at (4,0){x:i};
    \node[rectangle, thick, rounded corners=7, draw=black, minimum size=5mm, double, double distance=1pt, fill=cyan, label=below:\tiny\tiny2](a4) at (6,0){y:i};
    \draw[-, line width=1.2pt] (a1) edge[red] node[above, color = black]{z} (a2) (a3) edge[dashed] node[above, color = black]{z} (a4);
    \draw (6.95,1) -- (6.95,-0.5);
    \end{tikzpicture}
    \hspace{1em}
    \begin{tikzpicture}
    \tikzstyle{every node}=[font=\ttfamily]
    \draw (1.3,0.75) node[align=left] {back(x,y,z:list)};
    \node[rectangle, thick, rounded corners=7, draw=black, minimum size=5mm, fill=cyan, label=below:\tiny\tiny1](a1) at (0,0){x};
    \node[rectangle, thick, rounded corners=7, draw=black, minimum size=5mm, double, double distance=1pt, fill=cyan, label=below:\tiny\tiny2](a2) at (2,0){y};
    \draw (3,0) node[] {$\Rightarrow$};
    \node[rectangle, thick, rounded corners=7, draw=black, minimum size=5mm, double, double distance=1pt, fill=cyan, label=below:\tiny\tiny1](a3) at (4,0){x};
    \node[rectangle, thick, rounded corners=7, draw=black, minimum size=5mm, fill=cyan, label=below:\tiny\tiny2](a4) at (6,0){y};
    \draw[-, line width=1.2pt] (a1) edge[dashed] node[above, color = black]{z} (a2) (a3) edge[cyan] node[above, color = black]{z} (a4);
    \end{tikzpicture}
    
    \end{mdframed}
    \caption{The program \texttt{component-numbering}.}
    \label{fig:count-con-fig}
    \end{figure}

\subsection{Program}

The program \texttt{component-numbering} works by first evoking the rule \texttt{init}, which marks an arbitrarily-labelled node grey, roots it and appends \texttt{1} (first component identifier) to its list label. If the rule fails to match, the graph is empty and the program terminates, given that \texttt{unroot} fails. The procedure \texttt{DFS} works analogously to that of \texttt{is-dag}, except it is undirected. The rule \texttt{move} propagates the identifier.

The first looping body \texttt{DFS!} propagates the numbering \texttt{1} in a single connected component of the graph. The next looping procedure repeats the process, except it invokes \texttt{next} instead of \texttt{init}. The rule \texttt{next} simply unroots the current rooted node, roots another unvisited (grey-marked) node and appends the identifier of the previous rooted node, incremented by $1$. Once all unvisited nodes are exhausted, the rule \texttt{next} fails to match and the \texttt{break} command is called. The rule \texttt{unroot} unroots the sole root of the graph.

\subsection{Time Complexity}

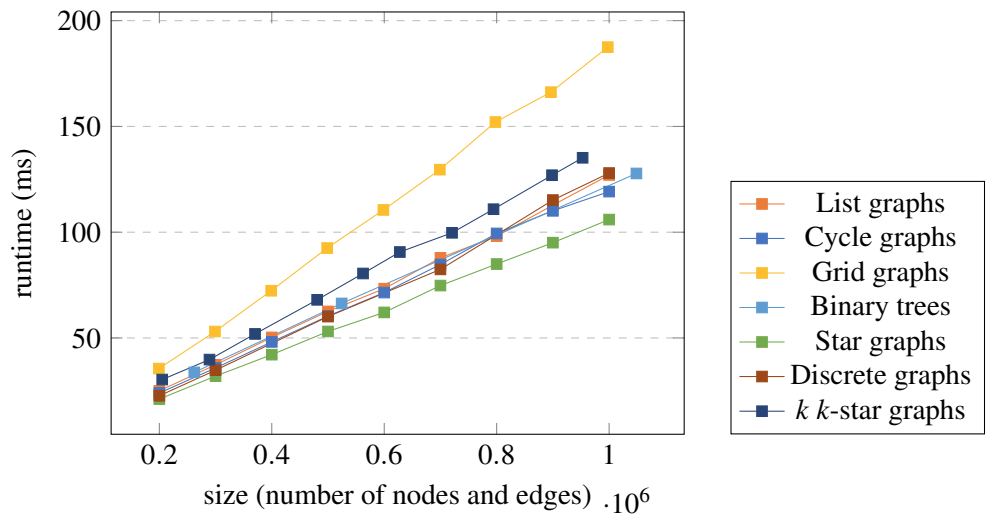
\begin{figure}[!ht]
    \centering
    \begin{tikzpicture}
    \begin{axis}[
      xlabel=size (number of nodes and edges),
      ylabel=runtime (ms), ylabel style={above=0.2mm},
      width=9.2cm,height=7.2cm,
      legend style={at={(1.525,0.6)}},
      ymajorgrids=true,
      grid style=dashed]
      \addplot[color=plot2, mark=square*] table [y=time, x=n]{Figures/Benchmarks/COUNT/count-list.dat};
      \addlegendentry{List graphs}
      \addplot[color=plot1, mark=square*] table [y=time, x=n]{Figures/Benchmarks/COUNT/count-cycle.dat};
      \addlegendentry{Cycle graphs}
      \addplot[color=plot4, mark=square*] table [y=time, x=n]{Figures/Benchmarks/COUNT/count-grid.dat};
      \addlegendentry{Grid graphs}
      \addplot[color=plot5, mark=square*] table [y=time, x=n]{Figures/Benchmarks/COUNT/count-tree.dat};
      \addlegendentry{Binary trees}
      \addplot[color=plot6, mark=square*] table [y=time, x=n]{Figures/Benchmarks/COUNT/count-star.dat};
      \addlegendentry{Star graphs}
      \addplot[color=plot7, mark=square*] table [y=time, x=n]{Figures/Benchmarks/COUNT/count-discrete.dat};
      \addlegendentry{Discrete graphs}
      \addplot[color=plot8, mark=square*] table [y=time, x=n]{Figures/Benchmarks/COUNT/count-k-star.dat};
      \addlegendentry{$k$ $k$-star graphs}
    \end{axis}  
    \end{tikzpicture}
    \caption{Measured performance of the program \texttt{component-numbering} under the modified compiler.}
    \label{fig:bench-count}
\end{figure}

The time complexity of the program \texttt{component-numbering} is linear. That is largely attributed to the fact that \texttt{init} and \texttt{next} match in constant time, which would have not been possible under the unmodified compiler. Figure \ref{fig:bench-count} offers corroborative evidence. As expected, the program exhibits a linear runtime on discrete graphs, which are disconnected and require $n-1$ calls of the rule \texttt{next}, with $n$ being the number of nodes.

\section{Case Study: Breadth-First Search}
The problem of traversing a graph in a breadth-first search (BFS) fashion in \gp{} is interesting, as previous techniques used to traverse graphs non-destructively in linear time involved variations of a depth-first search. Bak proposed, in his PhD thesis \cite{bak2015gp}, an implementation of the BFS algorithm in \gp{}. However, that program ran in quadratic time. That is due to there being no way, prior to the compiler enhancement of this paper, to find a node to expand from in constant time, given that the next node to expand from is not necessarily adjacent to the one being expanded during a BFS.

\definecolor{gp2pink}{RGB}{255, 153, 238}

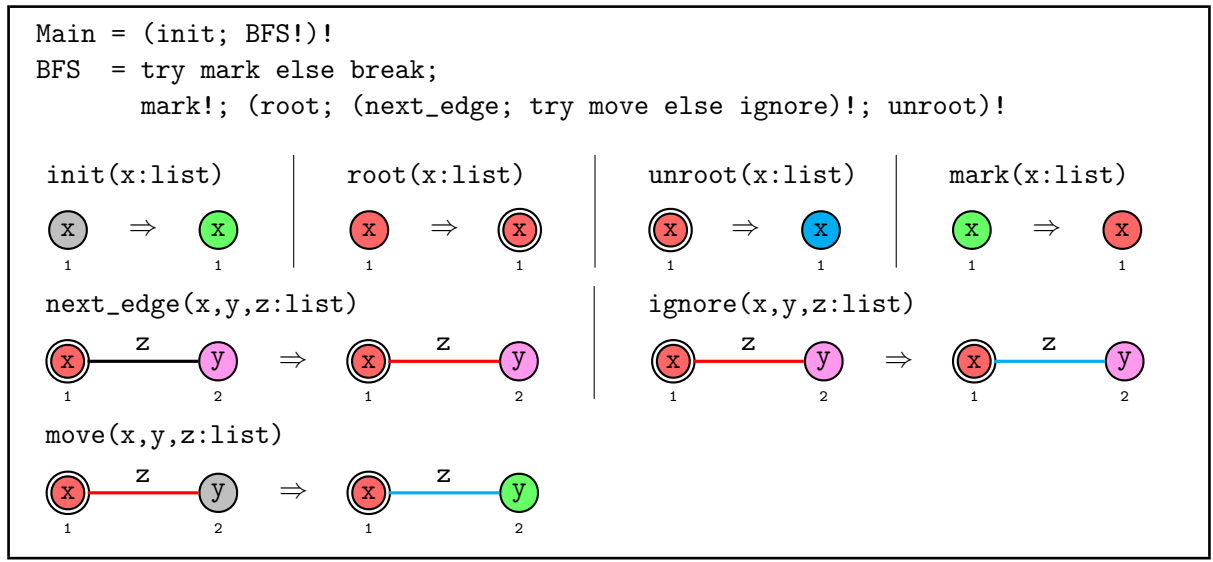
\begin{figure}[!ht]
    \begin{mdframed}[linewidth=1pt]
    \begin{verbatim}
Main = (init; BFS!)!
BFS  = try mark else break; 
       mark!; (root; (next_edge; try move else ignore)!; unroot)!
    \end{verbatim}
    \begin{tikzpicture}
    \tikzstyle{every node}=[font=\ttfamily]
    \draw (0.9,0.75) node[align=left] {init(x:list)};
    \node[rectangle, thick, rounded corners=7, draw=black, minimum size=5mm, fill=gray!50, label=below:\tiny\tiny1](a2) at (0,0){x};
    \draw (1,0) node[] {$\Rightarrow$};
    \node[rectangle, thick, rounded corners=7, draw=black, minimum size=5mm, fill=green!60, label=below:\tiny\tiny 1](a2) at (2,0){x};
    \draw[-, thick];
    \draw (3,1) -- (3,-0.5);
    \end{tikzpicture}
    \hspace{1em}
    \begin{tikzpicture}
    \tikzstyle{every node}=[font=\ttfamily]
    \draw (0.9,0.75) node[align=left] {root(x:list)};
    \node[rectangle, thick, thick, rounded corners=7, draw=black,  minimum size=5mm, fill=red!60, label=below:\tiny\tiny 1](a2) at (0,0){x};
    \draw (1,0) node[] {$\Rightarrow$};
    \node[rectangle, thick, rounded corners=7, draw=black, double, double distance=1pt, minimum size=5mm,  fill=red!60, label=below:\tiny\tiny1](a2) at (2,0){x};
    \draw[-, thick];
    \draw (3,1) -- (3,-0.5);
    \end{tikzpicture}
    \hspace{1em}
    \begin{tikzpicture}
    \tikzstyle{every node}=[font=\ttfamily]
    \draw (1.1,0.75) node[align=left] {unroot(x:list)};
    \node[rectangle, thick, rounded corners=7, draw=black, double, double distance=1pt, minimum size=5mm, fill=red!60, label=below:\tiny\tiny1](a2) at (0,0){x};
    \draw (1,0) node[] {$\Rightarrow$};
    \node[rectangle, thick, rounded corners=7, draw=black, minimum size=5mm, fill=cyan, label=below:\tiny\tiny 1](a2) at (2,0){x};
    \draw[-, thick];
    \draw (3,1) -- (3,-0.5);
    \end{tikzpicture}
    \hspace{1em}
    \begin{tikzpicture}
    \tikzstyle{every node}=[font=\ttfamily]
    \draw (0.9,0.75) node[align=left] {mark(x:list)};
    \node[rectangle, thick, rounded corners=7, draw=black, minimum size=5mm, fill=green!60, label=below:\tiny\tiny1](a2) at (0,0){x};
    \draw (1,0) node[] {$\Rightarrow$};
    \node[rectangle, thick, rounded corners=7, draw=black, minimum size=5mm, fill=red!60, label=below:\tiny\tiny 1](a2) at (2,0){x};
    \draw[-, thick];
    \end{tikzpicture}
    
    \begin{tikzpicture}
    \tikzstyle{every node}=[font=\ttfamily]
    \draw (1.8,0.75) node[align=left] {next\_edge(x,y,z:list)};
    \node[rectangle, thick, rounded corners=7, draw=black, double, double distance=1pt, minimum size=5mm, fill=red!60, label=below:\tiny\tiny1](a1) at (0,0){x};
    \node[rectangle, thick, rounded corners=7, draw=black, minimum size=5mm, fill=gp2pink, label=below:\tiny\tiny2](a2) at (2,0){y};
    \draw (3,0) node[] {$\Rightarrow$};
    \node[rectangle, thick, rounded corners=7, draw=black, double, double distance=1pt, minimum size=5mm,fill=red!60, label=below:\tiny\tiny1](a3) at (4,0){x};
    \node[rectangle, thick, rounded corners=7, draw=black, minimum size=5mm, fill=gp2pink, label=below:\tiny\tiny2](a4) at (6,0){y};
    \draw[-, line width=1.2pt] (a1) edge[black] node[above, color = black]{z} (a2) (a3) edge[red] node[above, color = black]{z} (a4);
    \draw (7,1) -- (7,-0.5);
    \end{tikzpicture}
    \hspace{1em}
    \begin{tikzpicture}
    \tikzstyle{every node}=[font=\ttfamily]
    \draw (1.45,0.75) node[align=left] {ignore(x,y,z:list)};
    \node[rectangle, thick, rounded corners=7, draw=black, double, double distance=1pt, minimum size=5mm, fill=red!60, label=below:\tiny\tiny1](a1) at (0,0){x};
    \node[rectangle, thick, rounded corners=7, draw=black, minimum size=5mm, fill=gp2pink, label=below:\tiny\tiny2](a2) at (2,0){y};
    \draw (3,0) node[] {$\Rightarrow$};
    \node[rectangle, thick, rounded corners=7, draw=black, double, double distance=1pt, minimum size=5mm,fill=red!60, label=below:\tiny\tiny1](a3) at (4,0){x};
    \node[rectangle, thick, rounded corners=7, draw=black, minimum size=5mm, fill=gp2pink, label=below:\tiny\tiny2](a4) at (6,0){y};
    \draw[-, line width=1.2pt] (a1) edge[red] node[above, color = black]{z} (a2) (a3) edge[cyan] node[above, color = black]{z} (a4);
    \end{tikzpicture}
    
    \begin{tikzpicture}
    \tikzstyle{every node}=[font=\ttfamily]
    \draw (1.3,0.75) node[align=left] {move(x,y,z:list)};
    \node[rectangle, thick, rounded corners=7, draw=black, double, double distance=1pt, minimum size=5mm, fill=red!60, label=below:\tiny\tiny1](a1) at (0,0){x};
    \node[rectangle, thick, rounded corners=7, draw=black, minimum size=5mm, fill=gray!50, label=below:\tiny\tiny2](a2) at (2,0){y};
    \draw (3,0) node[] {$\Rightarrow$};
    \node[rectangle, thick, rounded corners=7, draw=black, double, double distance=1pt, minimum size=5mm, fill=red!60, label=below:\tiny\tiny1](a3) at (4,0){x};
    \node[rectangle, thick, rounded corners=7, draw=black, minimum size=5mm,fill=green!60, label=below:\tiny\tiny2](a4) at (6,0){y};
    \draw[-, line width=1.2pt] (a1) edge[red] node[above, color = black]{z} (a2) (a3) edge[cyan] node[above, color = black]{z} (a4);
    \end{tikzpicture}
    
    \end{mdframed}
    \caption{The program \texttt{bfs}.}
    \label{fig:bfs-fig}
    \end{figure}

In this section, we present the program \texttt{bfs}, capable of carrying out a breadth-first search of a graph in linear time with respect to its size and the following specification.

\begin{description}
\item[\textbf{Input:}] An arbitrary \gp{} host graph such that
\begin{enumerate}
        \item each node is non-rooted and marked grey, and
        \item each edge is unmarked.
    \end{enumerate}
\item[\textbf{Output:}] A host graph structually isomorphic to the input graph where all nodes and edges are marked blue.
\end{description}

\subsection{Program}

The program exploits the advantages of the compiler modifications in order to find the next node to expand from in constant time. The program \texttt{bfs} consists of a loop, \texttt{(init; BFS!)}, which itself contains a procedure, \texttt{BFS}, which marks all green nodes in the host graph red and, subsequently, marks all grey nodes directedly adjacent to the red nodes green. Once a red node has been expanded from, it is marked blue. That procedure loops as long as the connected component of the node marked green by \texttt{init} contains non-blue nodes.

Intuitively, at the beginning of the execution of \texttt{BFS}, the program seeks to mark all nodes directedly adjacent to the node marked green by \texttt{init}. Firstly, the rule \texttt{mark} is applied as long as possible to mark all green-marked nodes red. Then, for as long as possible, the program picks some red node (which was previously green) with the rule \texttt{root}, and expands from it as long as possible. The nested looping procedure ends when \texttt{next\_edge} is no longer applicable, implying that there is no expansion left from the node picked. The rule \texttt{unroot} then marks the chosen node blue, indicating that it was fully processed, and moves on to the next red-marked node. The upper parenthetical looping procedure within \texttt{BFS!} terminates when \texttt{root} no longer applies, indicating that all nodes that were previously green at the beginning of \texttt{BFS} have been processed. Once \texttt{BFS!} terminates, the rule \texttt{init} is called again to start a bread-first search procedure from a different non-visited connected component. The loop continues until all nodes in the host graphs are visited.

\subsection{Time Complexity}

The program \texttt{bfs} runs in linear time with respect to the size of the graph (i.e. the number of nodes and edges).

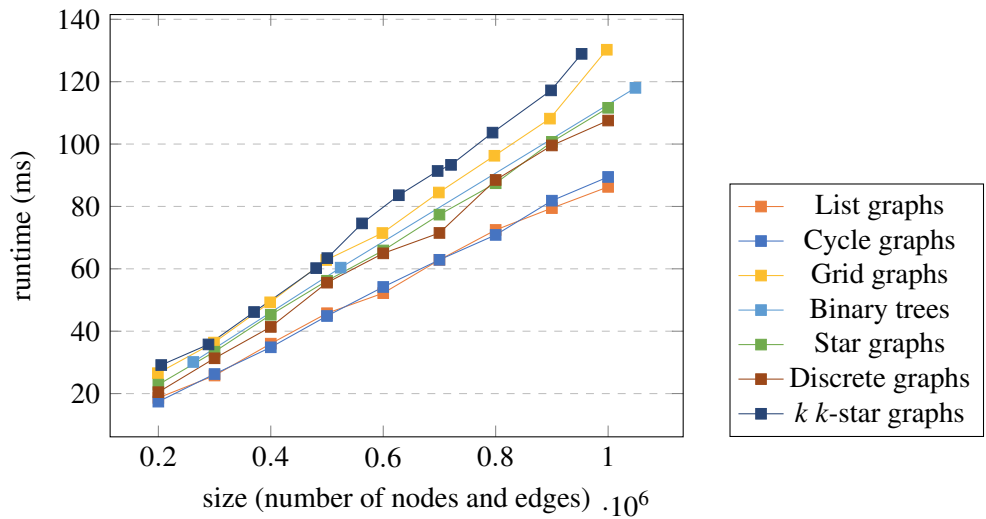
\begin{figure}[!ht]
    \centering
    \begin{tikzpicture}
    \begin{axis}[
      xlabel=size (number of nodes and edges),
      ylabel=runtime (ms), ylabel style={above=0.2mm},
      width=9.2cm,height=7.2cm,
      legend style={at={(1.525,0.6)}},
      ymajorgrids=true,
      grid style=dashed]
      \addplot[color=plot2, mark=square*] table [y=time, x=n]{Figures/Benchmarks/BFS/bfs-list.dat};
      \addlegendentry{List graphs}
      \addplot[color=plot1, mark=square*] table [y=time, x=n]{Figures/Benchmarks/BFS/bfs-cycle.dat};
      \addlegendentry{Cycle graphs}
      \addplot[color=plot4, mark=square*] table [y=time, x=n]{Figures/Benchmarks/BFS/bfs-grid.dat};
      \addlegendentry{Grid graphs}
      \addplot[color=plot5, mark=square*] table [y=time, x=n]{Figures/Benchmarks/BFS/bfs-tree.dat};
      \addlegendentry{Binary trees}
      \addplot[color=plot6, mark=square*] table [y=time, x=n]{Figures/Benchmarks/BFS/bfs-star.dat};
      \addlegendentry{Star graphs}
      \addplot[color=plot7, mark=square*] table [y=time, x=n]{Figures/Benchmarks/BFS/bfs-discrete.dat};
      \addlegendentry{Discrete graphs}
      \addplot[color=plot8, mark=square*] table [y=time, x=n]{Figures/Benchmarks/BFS/bfs-k-star.dat};
      \addlegendentry{$k$ $k$-star graphs}
    \end{axis}  
    \end{tikzpicture}
    \caption{Measured performance of the program \texttt{bfs} under the modified compiler.}
    \label{fig:bench-bfs}
\end{figure}

\section{Conclusion}
We have demonstrated by case studies how to implement in \gp{} graph algorithms based on depth-first and breadth-first search such that a linear runtime is achieved, even if input graphs have an unbounded node degree or are possibly disconnected. Addressing the issues of unbounded degree and disconnectedness has been an open problem since the publication of the first paper on rooted graph transformation \cite{bak2012rooted}. Up to now, only certain graph reduction programs that destroy their input graphs could be designed to run in linear time on graph classes that have an unbounded node degree or contain disconnected graphs \cite{campbell2022fast}.

Our approach involves both enhancing the graph data structure generated by the \gp{} compiler and developing a programming technique that leverages the new graph representation. Previously, the graph data structure in the C program generated by the compiler stored all host-graph nodes in a single linked list. Hence, if host graph nodes may have different marks, searches within this list required linear time, preventing constant-time rule matching. In contrast, the new data structure finds a node with a given mark or an edge with a given mark and orientation in constant time.

We speculate that all linear-time graph algorithms based on depth-first or breadth-first search can be implemented as \gp{} programs running in linear time. More generally, we intend to implement a \gp{} library of advanced data structures, such as priority queues, Fibonacci heaps, or AVL trees, to support programmers in constructing \gp{} versions of conventional graph algorithms that match the time complexity achievable in the imperative setting.


\bibliographystyle{eptcs}
\bibliography{generic}
\nocite{*}
\end{document}